\documentclass[journal,10pt]{IEEEtran}
\usepackage{kotex}
\usepackage{cite}
\usepackage{lettrine}
\usepackage{graphicx}
\usepackage{setspace}
\usepackage{xcolor}
\usepackage{amsmath}
\usepackage{dsfont}
\usepackage{amsfonts}
\usepackage{lipsum} 
\usepackage{multicol}
\usepackage{float}
\usepackage{bm}
\usepackage{color}
\usepackage{etoolbox}
\usepackage{soul}
\usepackage{amssymb}
\usepackage{algorithm}
\usepackage{algorithmic}
\usepackage{mathtools,amssymb} 
\usepackage{float}
\usepackage{amsthm}
\usepackage[caption = false,subrefformat=parens,labelformat=parens]{subfig}
\usepackage{adjustbox}
\usepackage{afterpage}
\usepackage{multirow}
\usepackage{mathrsfs}
\usepackage{hyperref}
\makeatletter

\newcommand{\E}{\operatorname*{\mathbb{E}}\ilimits@}

\makeatother

\begin{document}

\title{Energy-Efficient Secure Communications via \\Joint Optimization of UAV Trajectory and \\Movable-Antenna Array Beamforming}

\author{Sanghyeok Kim,~\IEEEmembership{Student Member,~IEEE,} Jinu Gong, and Joonhyuk Kang,~\IEEEmembership{Member,~IEEE}

\thanks{S. Kim and J. Kang are with the School of Electrical Engineering, Korea Advanced Institute of Science and Technology, Daejeon 34141, Korea (e-mail: starmoon13, jkang@kaist.ac.kr)}
\thanks{J. Gong is with the Department of Applied Artificial Intelligence, Hansung University, Seoul, Korea (e‑mail: jinugong@hansung.kr)}
\thanks{Manuscript received XXX, XX, 2025; revised XXX, XX, 2025.}}

\markboth{IEEE Wireless Communications Letters,~Vol.~XX, No.~XX, XXX~2025}
{}

\maketitle

\begin{abstract}
This paper investigates the potential of unmanned aerial vehicles (UAVs) equipped with movable-antenna (MA) arrays to strengthen security in wireless communication systems. We propose a novel framework that jointly optimizes the UAV trajectory and the reconfigurable beamforming of the MA array to maximize secrecy energy efficiency, while ensuring reliable communication with legitimate users. By exploiting the spatial degrees of freedom enabled by the MA array, the system can form highly directional beams and deep nulls, thereby significantly improving physical layer security. Numerical results demonstrate that the proposed approach achieves superior secrecy energy efficiency, attributed to the enhanced spatial flexibility provided by the movable antenna architecture.
\end{abstract}


\IEEEpeerreviewmaketitle

\setlength{\abovedisplayskip}{3pt}
\setlength{\belowdisplayskip}{5pt}

\vspace{-0.2cm}
\section{Introduction}
\lettrine{P}{hysical} layer security (PLS) has emerged as a key enabler in modern wireless communications, offering provable security guarantees that remain robust even against adversaries with unlimited computational power \cite{bloch2011physical}. This paradigm has garnered increasing attention for its potential to complement conventional cryptographic methods and address emerging threats in next-generation wireless networks. Among various PLS techniques, jamming has been extensively studied as an effective strategy to combat eavesdropping. The central idea is to transmit artificial noise that impairs the eavesdropper’s reception while preserving reliable communication with legitimate users \cite{hong2020artificial}.

The effectiveness of jamming is highly influenced by the spatial distribution of the jamming signal. To maximize interference at the eavesdropper while minimizing disruption to legitimate users, it is critical to position the jammer in close proximity to potential eavesdroppers while avoiding interference to legitimate users. This spatial requirement has motivated the integration of unmanned aerial vehicles (UAVs) as mobile jamming platforms, as their mobility allows dynamic adjustment of their position relative to both legitimate users and eavesdroppers \cite{zhou2019uav,zhang2019securing}. However, UAVs are subject to strict energy constraints, which limit their operational duration and system performance. In conventional UAV-based jamming strategies, when the legitimate user and the eavesdropper are aligned from the jammer’s perspective, the UAV often takes a detour to avoid beamforming interference. Although this maneuver preserves the secrecy capacity, it incurs higher energy consumption due to the extended flight path \cite{zhang2019securing}.

To overcome these challenges while ensuring robust security performance, we propose the use of movable-antenna (MA) arrays mounted on UAVs. MA technology enables dynamic reconfiguration of antenna element positions, allowing the optimization of radiation patterns and spatial diversity—offering clear advantages over conventional fixed-antenna arrays \cite{shao20256dma, zhu2025tutorial}. When integrated with UAVs, the MA array produces an additional spatial degree of freedom, which can be exploited to enhance physical layer security while maintaining high energy efficiency.

Recently, several researches have explored combining UAV and MA technology. Studies in \cite{liu2024uav},\cite{shen2025ris} showed MA arrays' benefits for UAV-terrestrial communications, while \cite{ren20256d} investigated MA arrays in cellular-connected UAV systems. The 6DMA was proposed in \cite{shao20246d}, \cite{shao2024exploiting} for ground BS-UAV receiver applications. However, most work has focused on throughput maximization rather than security, and only consider the UAV as a transceiver. There has been limited research on physical layer security with UAVs and MAs, and furthermore, no prior work has investigated the use of MA-equipped UAV as jammer.

We propose a compact framework that integrates UAV mobility with movable-antenna (MA) array flexibility to utilize it as a favorable jammer and achieve optimal secrecy performance. The approach enables fine-grained spatial control of jamming signals: the UAV achieves tighter flight paths, while the MA array dynamically forms optimal beam patterns. This synergy between mobility and antenna adaptability significantly enhances physical layer security, enabling robust and energy-efficient secure wireless communications.

\vspace{-0.2cm}
\section{System Model and Problem Formulation}
\subsection{System Model}
We consider a UAV-assisted secure wireless network where a single rotary-wing UAV serves as a friendly jammer to enhance communication security. The system consists of an $N_{\text{B}}$-antennas base station (BS), a single-antenna legitimate user, and a single-antenna eavesdropper (eve). The UAV follows an optimized trajectory from $\mathbf{q}_{I}$ to $\mathbf{q}_{F}$ within flight time $T_{\text{flight}}$ divided into $N_{\text{step}}$ time slots.

The BS is located at $\mathbf{q}_{\text{B}}=[q_{\text{B},x},q_{\text{B},y},H_{\text{B}}]$. The UAV's movement is constrained by velocity limits $v_{\min} \leq \|\mathbf{v}_{\text{J}}[n]\| \leq v_{\max}$ and maximum acceleration $\|\mathbf{a}_{\text{J}}[n]\| \leq a_{\max}$. The UAV maintains a fixed altitude $H_{\text{J}}$, with its location at time slot $n$ denoted by $\mathbf{q}_{\text{J}}[n]=[q_{\text{J},x}[n],q_{\text{J},y}[n],H_{\text{J}}]$.

The legitimate user and eve locations are denoted as $\mathbf{q}_{\text{i}}=[q_{\text{i},x},q_{\text{i},y},0]$, where $i\in\lbrace u, e\rbrace$. Through secure coordination protocols, it is assumed that the legitimate user's exact position $\mathbf{q}_{u}$ is known to both BS and UAV, while for eve, only an approximate location $\tilde{\mathbf{q}}_{e}$ is available \cite{zhong2018secure}. The uncertainty in eve's true position is bounded by region $\Theta\triangleq\lbrace\mathbf{q}: \lVert\mathbf{q}-\tilde{\mathbf{q}}_{e}\rVert\leq\epsilon\rbrace$, where $\epsilon$ represents the maximum localization error.

\subsubsection{MA Array and Local Coordinates \texorpdfstring{\cite{shao20256dma}}{[shao20256dma]}}
The UAV carries a MA array with $N_{\text{MA}}=N_x N_y$ elements on a rectangular panel of dimensions $l_x$ and $l_y$. The array is mounted at the bottom of the UAV, requiring a local coordinate system where the $Z'$-axis points downward and the $Y'$-axis is inverted.

The antenna array's spatial orientation can be dynamically adjusted through rotations about three perpendicular axes $X', Y', Z'$. At each time slot $n$, the array's orientation is specified by the rotation vector $\bm{\phi}[n] = [\phi_{X'}[n], \phi_{Y'}[n], \phi_{Z'}[n]]^T$. The transformed local coordinate system can be obtained through a composite transformation:
\begin{equation} \small
U[n] = R_{Z'}[n]R_{Y'}[n]R_{X'}[n]\begin{bmatrix}1 & 0 & 0 \\ 0 & -1 & 0 \\ 0 & 0 & -1\end{bmatrix},
\end{equation}
where $R_{X'}[n]$, $R_{Y'}[n]$, and $R_{Z'}[n]$ represent the rotation matrices that transform the initial array configuration around the local coordinate axes by rotation angles $\phi_{X'}[n]$, $\phi_{Y'}[n]$, and $\phi_{Z'}[n]$, respectively.

\subsubsection{Array Response and Channel Vector}
To characterize the spatial relationship between the array and ground nodes, we first transform the direction vector from global to local coordinates. For any transmitting node $t \in \{\text{B}, \text{J}\}$ and receiving node $r \in \{u,e\}$, this transformation at time slot $n$ is given by
\begin{equation} \small
\mathbf{u}_{t,r}[n] = U[n]\frac{\mathbf{q}_r - \mathbf{q}_{t}[n]}{\|\mathbf{q}_r - \mathbf{q}_{t}[n]\|}.
\end{equation}

Specifically, since the $(a,b)$-th element's position in the MA array at time slot $n$ can be represented by $\mathbf{p}_{a,b} = [(a-1)\lambda/2, (b-1)\lambda/2, 0]^T$ in the local coordinate system, the array response vector $\mathbf{g}_{t,r}[n] \in \mathbb{C}^{N_{\text{MA}}}$ for the link between nodes $t$ and $r$ at time slot $n$ can then be expressed as:
\begin{equation} \small
[\mathbf{g}_{t,r}[n]]_{(a-1)N_y + b} = e^{j\frac{2\pi f}{c}(\mathbf{u}_{t,r}[n]^T\mathbf{p}_{a,b})}.
\end{equation}

Given the UAV's elevated position and high-frequency bands, the aerial-ground links follow line-of-sight (LoS) propagation \cite{zhang2019securing}. This characteristic allows us to exclude multipath propagation effects from ground reflections \cite{ren20256d}. Consequently, for any link between a transmitting node $t \in \{\text{B}, \text{J}\}$ and a receiving node $r \in \{\text{u}, \text{e}\}$, the channel vector is modeled as:
\begin{equation} \small
\mathbf{h}_{t,r}[n] = \sqrt{\beta_{t,r}[n]}\mathbf{g}_{t,r}[n],
\end{equation}
where $\beta_{t,r}[n] = (\frac{c}{4\pi f})^2 d_{t,r}[n]^{-\alpha_{t,r}}$ represents the large-scale path loss with $d_{t,r}[n] = \lVert \mathbf{q}_t[n] - \mathbf{q}_r \rVert$ being the distance between nodes and $\alpha_{t,r}$ denoting the path loss exponent.

\subsection{Problem Formulation}
\vspace{-0.1cm}
In this paper, we focus on maximizing the sum secrecy energy efficiency (SEE) of the system by jointly optimizing the UAV trajectory $\textbf{q}_{\text{J}}[n]$, MA orientation angles $\bm{\phi}_{\text{J}}[n]$, and MA beamforming vector $\textbf{w}_{\text{J}}[n]$. As a first step, we define two key terms that will be essential to our formulation.

\subsubsection{Sum Secrecy Capacity}
Let $\mathbf{w}_{\text{B}}[n]\in\mathbb{C}^{N_{\text{B}}\times1}$ and $\mathbf{w}_{\text{J}}[n]\in\mathbb{C}^{N_{\text{MA}}\times1}$ be the unit-bounded transmit beamforming vectors of the BS and MA at time slot $n$, respectively. The received signal at node $i \in \{u,e\}$ is given by:
\begin{equation} \small
y_i[n] = \mathbf{h}_{\text{B},i}^H[n]\mathbf{w}_{\text{B}}[n]\sqrt{P_{\text{B}}}s_{\text{B}} + \mathbf{h}_{\text{J},i}^H[n]\mathbf{w}_{\text{J}}[n]\sqrt{P_{\text{J}}}s_{\text{J}} + z_i[n],
\end{equation}
where $P_{\text{B}}$ and $P_{\text{J}}$ denote the transmit power of the BS and UAV, respectively; $s_{\text{B}}$ and $s_{\text{J}}$ are the normalized transmitted symbols; and $z_i[n] \sim \mathcal{C}\mathcal{N}(0, \sigma_i^2)$ represents the additive white Gaussian noise (AWGN) at node $i$ with average power $\sigma_i^2$.
\vspace{-0.1cm}
The received signal-to-interference-plus-noise ratio (SINR) at node $i \in \{u,e\}$ is then expressed as
\begin{equation} \small
\gamma_i[n] = \frac{P_{\text{B}}|\mathbf{h}_{\text{B},i}^H[n]\mathbf{w}_{\text{B}}[n]|^2}{P_{\text{J}}|\mathbf{h}_{\text{J},i}^H[n]\mathbf{w}_{\text{J}}[n]|^2 + \sigma_i^2}.
\end{equation}

Considering the partial knowledge of the eve's position, we analyze the worst-case secrecy capacity by evaluating the rate differential between the legitimate user's and that of the eve:
\begin{equation} \small
r_{\text{sec}}[n] = \left[\log_2(1+\gamma_u[n]) - \max_{\mathbf{q}_{e}\in\Theta}\log_2(1+\gamma_e[n])\right]^+,
\end{equation}
where $[x]^+ = \max(0,x)$. 
To effectively address the optimization problem afterwards, we propose a tractable solution approach. The main challenge lies in handling the term $\max_{\mathbf{q}_e \in \Theta} \log_2(1+\gamma_e[n])$. We can address this challenge by deriving explicit bounds on the channel gains. For the BS-eve link, the maximum channel gain is achieved by $\tilde{\mathbf{h}}_{\text{B},e}[n] = \frac{\beta_0}{(\|\mathbf{q}_{\text{B}} - \tilde{\mathbf{q}}_e\| - \epsilon)^2}$ with $\beta_0$ denoting a path loss at 1$m$, when the eve is optimally positioned within its uncertainty region. Similarly, the minimum interference gain for the UAV-eve link can be obtained as $\tilde{\mathbf{h}}_{\text{J},e}[n] = \frac{\beta_0}{(\|\mathbf{q}_{\text{J}}[n] - \tilde{\mathbf{q}}_e\| + \epsilon)^2}$. Based on these bounds, we can establish an upper bound for the eve's achievable rate:
\begin{equation} \small
\max_{\mathbf{q}_e \in \Theta} \log_2(1+\gamma_e[n]) \leq \tilde{r}_e[n] = \log_2 \left(1 + \frac{P_{\text{B}}\tilde{\mathbf{h}}_{\text{B},e}[n]}{P_{\text{J}}\tilde{\mathbf{h}}_{\text{J},e}[n] + \sigma_e^2}\right).
\end{equation}

This enables us to derive a tractable lower bound on the instantaneous secrecy capacity: $\tilde{r}_{\text{sec}}[n] = [\log_2(1+\gamma_u[n]) - \tilde{r}_e[n]]^+$. Subsequently, the sum secrecy capacity over $N_{\text{step}}$ time slots can be expressed as:
\begin{equation} \label{eq:sec_rate} \small
R_{\text{sec}} = \sum\limits_{n=1}^{N_{\text{step}}} \tilde{r}_{\text{sec}}[n].
\end{equation}

\subsubsection{Energy Consumption Model}
The total energy consumption is consists of propulsion power $P_{\text{UAV}}$, MA movement power $P_{\text{MA}}$, and communication power $P_{\text{com}}$, and their respective operation times. We consider a UAV capable of hovering and moving at constant velocity (i.e., $v_{\text{min}}=a_{\text{max}}=0$). Under this assumption, the propulsion power required for the UAV flight can be described as \cite{yang2019energy}:
\begin{equation} \small
\begin{aligned}
P_{\text{UAV}}[n] &= P_0 \left( 1 + \frac{3v[n]^2}{U_{\text{tip}}^2} \right) + P_1 \left( \sqrt{1 + \frac{v[n]^4}{4v_0^4}} - \frac{v[n]^2}{2v_0^2} \right)^{\frac{1}{2}} \\
& \qquad\qquad\qquad\qquad\qquad\qquad\quad + \frac{1}{2} r_{\text{drag}}\rho S A v[n]^3,
\end{aligned}
\end{equation}
where $v[n] := \Delta_{\mathbf{q}_J}[n]/\Delta t$ with $\Delta_{\mathbf{q}_J}[n]$ and $\Delta t$ denoting the UAV moving distance at time slot $n$ and time duration for each slot, respectively, and $P_0, P_1, U_{\text{tip}}^2, v_0, r_{\text{drag}}, \rho, S, A$ are the propulsion energy consumption model parameters for the rotary-wing UAV \cite{yang2019energy}.

The transmission power consumption is given by $P_{\text{com}}[n] = P_{\text{J}}\|\mathbf{w}_{\text{J}}[n]\|^2$, and power consumption of the MA is defined as
\begin{equation} \small
P_{\text{MA}}[n] = P_{\text{base}} + \zeta \cdot \Delta \phi_{X'}[n] + \xi \cdot \Delta \phi_{Z'}[n],
\end{equation}
where $P_{\text{base}}, \zeta, \xi$ are the MA power consumption model parameters. The MA operation time $t_{\text{MA}}[n]$ is determined by $\Delta\phi_{X'}[n]$, $\Delta\phi_{Z'}[n]$, and the maximum angular velocities for elevation and azimuth adjustments \cite{bai2024movable}.

Considering the aforementioned power consumption terms and their respective operation times, the total energy consumption for a time step $n$ can be expressed as 
\begin{equation} \small
E_{\text{total}}[n] = P_{\text{UAV}}[n]\Delta t + P_{\text{MA}}[n]\cdot t_{\text{MA}}[n] + P_{\text{com}}[n]\Delta t.
\end{equation}

Accordingly, the sum secrecy energy efficiency (SEE) maximization problem can be formulated as:
\begin{subequations} \small
\begin{align}
    &\max_{\substack{\lbrace\textbf{q}_{\text{J}}[n], \bm{\phi}_{\text{J}}[n], \textbf{w}_{\text{J}}[n]\rbrace}} &&\frac{\sum\limits_{n=1}^{N_{\text{step}}} \tilde{r}_{\text{sec}}[n]\Delta t}{\sum_{n=1}^{N_{\text{step}}} E_{\text{total}}[n]}, \tag{13} \label{eq:obj} \\
    &\qquad\text{s.t.} &&\lbrace\textbf{q}_{\text{J}}[0], \textbf{q}_{\text{J}}[N_{\text{step}}]\rbrace = \lbrace\textbf{q}_{I}, \textbf{q}_{F}\rbrace, \label{eq:const1} \\
    &&&\lVert\textbf{q}_{\text{J}}[n]-\textbf{q}_{\text{J}}[n-1]\rVert \leq V_{\text{max}}\Delta t, \label{eq:const2} \\
    &&&\mathbf{\phi}_{X'}[n],\mathbf{\phi}_{Z'}[n] \in \left[-\frac{\pi}{2},\,\frac{\pi}{2}\right], \label{eq:const3} \\
    &&&\lVert\textbf{w}_{\text{J}}[n]\rVert \leq 1, \label{eq:const4}
\end{align}
\end{subequations}
where the constraint \eqref{eq:const2} limits the UAV moving distance within one time slot, and the constraint \eqref{eq:const3} ensures that the MA orientation angles remain within their physical limits.

\section{Proposed Solution}
It is worth noting that, according to the theoretical analysis presented in \cite{zhang2019securing}, the optimization always leads to a non-negative objective value. Thus, we can simplify our problem by removing the non-negative operator $\lbrack\cdot\rbrack^{+}$ from the objective function for the secrecy function maximization. As such, the optimization problem can be formulated as:
\begin{equation} \label{eq:SEE} \small
    \max_{\lbrace\textbf{q}_{\text{J}}[n], \bm{\phi}_{\text{J}}[n], \textbf{w}_{\text{J}}[n]\rbrace} \quad \bar{\eta}_{\text{SEE}}, \quad \text{s.t.} \quad \eqref{eq:const1}, \eqref{eq:const2}, \eqref{eq:const3} \text{ and } \eqref{eq:const4},
\end{equation}
where $\bar{\eta}_{\text{SEE}}$ denotes the objective function in \eqref{eq:obj} without the non-negative operator.

The formulated problem is non-convex due to the coupling between optimization variables and the fractional form of the objective function. To tackle this problem, we propose an alternating optimization (AO) approach that iteratively optimizes one set of variables while fixing the others.

\subsection{Optimizing Trajectory Given Beamforming Vector and Angles}
Now we optimize the UAV trajectory {$\mathbf{q}_{\text{J}}[n]$} under given beamforming vector {$\mathbf{w}_{\text{J}}[n]$} and MA angle {$\bm{\phi}_{\text{J}}[n]$}, for which the problem is formulated as:
\begin{equation} \label{eq:traj_sub_obj} \small
\hspace{-0.2cm}\max_{\lbrace\textbf{q}_{\text{J}}[n]\rbrace} \quad \frac{\sum\limits_{n=1}^{N_{\text{step}}} \bar{r}_{\text{sec}}[n]}{\sum\limits_{n=1}^{N_{\text{step}}} P_{\text{UAV}}[n]\Delta t + E_{\text{MA}} + E_{\text{com}}}, \quad \text{s.t.} \quad \eqref{eq:const1}, \eqref{eq:const2},
\end{equation}
where $\bar{r}_{\text{sec}}[n]$ denotes the secrecy capacity in \eqref{eq:sec_rate} without the non-negative operator.

To solve this problem, we first introduce auxiliary variables to modify the objective to a more tractable form. With $\lbrace\mu[n]\rbrace$, $\lbrace\nu[n]\rbrace$, $\lbrace\tau[n]\rbrace$, and $\lbrace\chi[n]\rbrace$, we can re-express the secrecy capacity and UAV energy consumption as:
\begin{equation} \label{eq:reform_secrecy} \small
\begin{aligned}
&\hspace{-0.2cm}\hat{r}_{\text{sec}}[n] =\\
&\hspace{-0.2cm} -\log_2 \left( \sigma_u^2 + \frac{\beta_0 P_{\text{J}}|\mathbf{g}_{\text{J},u}^H[n]\mathbf{w}_{\text{J}}[n]|^2}{\mu[n]} \right) \\
&\hspace{-0.2cm} + \log_2 \left( \sigma_e^2 + \beta_0 P_{\text{J}}|\mathbf{h}_{\text{J},e}^H[n]\mathbf{w}_{\text{J}}[n]|^2 \right) \\
&\hspace{-0.2cm} - \log_2 \left( \sigma_e^2 + \frac{\beta_0 P_{\text{B}}|\mathbf{g}_{\text{B},e}^H[n]\mathbf{w}_{\text{B}}[n]|^2}{\nu[n]} + \frac{\beta_0 P_{\text{J}}|\mathbf{g}_{\text{J},e}^H[n]\mathbf{w}_{\text{J}}[n]|^2}{\tau[n]} \right) \\
&\hspace{-0.2cm} + \log_2 \left( \sigma_u^2 + \beta_0 P_{\text{B}}|\mathbf{h}_{\text{B},u}^H[n]\mathbf{w}_{\text{B}}[n]|^2 + \beta_0 P_{\text{J}}|\mathbf{h}_{\text{J},u}^H[n]\mathbf{w}_{\text{J}}[n]|^2 \right),
\end{aligned}
\end{equation}
\begin{equation} \label{eq:reform_energy} \footnotesize
E_{\text{UAV}}[n] = \sum\limits_{n=1}^{N_{\text{step}}} \Delta t \left[P_0\left(1 + \frac{3v[n]^2}{U_{\text{tip}}^2}\right) + \frac{1}{2}r_{\text{drag}}\rho S A v[n]^3\right] + P_1\cdot\chi[n].
\end{equation}
By substituting these reformed expressions, the main problem can be rewritten as:
\begin{subequations} \label{eq:reform_main} \small
\begin{align}
&\max_{\substack{\lbrace\mathbf{q}_{\text{J}}[n],\\ \mu[n],\nu[n],\tau[n], \chi[n]\rbrace}} &&\frac{\sum_{n=1}^{N_{\text{step}}} \hat{r}_{\text{sec}}[n]}{E_{\text{UAV}}[n] + E_{\text{MA}} + E_{\text{com}}}, \tag{18} \label{eq:reform_obj}\\
&\qquad\quad\text{s.t.} &&e^{\frac{2\mu[n]}{\alpha_{\text{J,u}}}} \leq H^2 + \lVert \mathbf{q}_{\text{J}}[n] - \mathbf{q}_{\text{u}} \rVert^2, \label{eq:reform_const1} \\
&&&e^{\frac{2\nu[n]}{\alpha_{\text{B,e}}}} \leq H^2 + (\lVert \mathbf{q}_{\text{B}}[n] - \tilde{\mathbf{q}}_{\text{e}} \rVert -\epsilon)^2, \label{eq:reform_const2} \\
&&&e^{\frac{2\tau[n]}{\alpha_{\text{J,e}}}} \leq H^2 + (\lVert \mathbf{q}_{\text{J}}[n] - \tilde{\mathbf{q}}_{\text{e}} \rVert +\epsilon)^2, \label{eq:reform_const3} \\
&&&e^{\frac{2\mu[n]}{\alpha_{\text{J,u}}}}, e^{\frac{2\nu[n]}{\alpha_{\text{B,e}}}}, e^{\frac{2\tau[n]}{\alpha_{\text{J,e}}}} \geq H^2, \label{eq:reform_const4} \\
&&&\chi[n]^2 \geq \sqrt{\Delta t^4 + \frac{\Delta_{q_{\text{J}}}[n]^4}{4v_0^4}} - \frac{\Delta_{q_{\text{J}}}[n]^2}{2v_0^2}, \label{eq:reform_const5} \\
&&&\text{(\ref{eq:const1})} \text{ and } \text{(\ref{eq:const2})}. \notag
\end{align}
\end{subequations}

Note that the constraints in \eqref{eq:reform_const1}, \eqref{eq:reform_const3}, and \eqref{eq:reform_const5} are non-convex. Also, the second and fourth components of $\hat{r}_{\text{sec}}[n]$ demonstrate convex behavior with respect to {$\mathbf{q}_{\text{J}}[n]$}. Consequently, problem \eqref{eq:reform_main} belongs to the non-concave problems.

To tackle the non-concavity, we employ an iterative approach using successive convex approximation (SCA) to solve problem \eqref{eq:reform_main}. Specifically, at each iteration $m$$\geq1$, we linearize the non-concave components around a local trajectory point {$\mathbf{q}_{\text{J}}^{(m)}[n]$} using first-order Taylor series approximation \cite{zhong2018secure}. After that, we employ the Dinkelbach method which converts the fractional objective into a sequence of more tractable linear problems to solve this approximated problem efficiently \cite{schaible1976fractional}. These problems are then solved iteratively until convergence, which is guaranteed by the monotonic improvement property of the Dinkelbach algorithm.

\subsection{Optimizing MA Angles Given Beamforming Vector and Trajectory}
Next, we focus on optimizing the MA orientation angles with fixed $\mathbf{w}_{\text{J}}[n]$ and $\mathbf{q}_{\text{J}}[n]$. Note that it is enough to only care rotations around the $X'$ and $Z'$ axes (denoted by $\mathbf{\phi}_{X'}[n]$ and $\mathbf{\phi}_{Z'}[n]$), by considering the mechanical equivalent configuration from the 2D square uniform arrays. 

The optimization problem can be formulated as:
\begin{subequations} \label{eq:angle_sub_obj} \small
\begin{align}
\max_{\bm{\phi}_{\text{J}}[n]} \quad &\frac{\sum\limits_{n=1}^{N_{\text{step}}} \bar{r}_{\text{sec}}[n]\Delta t}{ E_{\text{UAV}} + \sum\limits_{n=1}^{N_{\text{step}}}P_{\text{MA}}[n]\cdot t_{\text{MA}}[n] + E_{\text{com}}}, \label{eq:angle_obj} \quad \text{s.t.} \quad \text{\eqref{eq:const3}}. \tag{19}
\end{align}
\end{subequations}

Due to the non-linear and fractional form objective function, applying conventional convex optimization techniques becomes intractable. Instead, we propose a feasible direction method to solve this problem.

Since \eqref{eq:angle_sub_obj} only imposes constraints on the angle region, we consider an efficient feasible direction method, i.e., the projected gradient descent method. Specifically, the gradient of the objective function in the $k$-th iteration is calculated as
\begin{equation}
\small
\nabla_{\phi_{i}}f(\bm{\phi}_{\text{J}}^{(k)}[n]) = \lim_{\epsilon_{\phi}\rightarrow0}\frac{f(\bm{\phi}_{\text{J}}^{(k)}[n]+\epsilon_{\phi}\cdot\mathbf{e}_{\phi_i})}{\epsilon_{\phi}}, \quad i = X',Z'
\end{equation}
where $f$ represents the $\bar{\eta}_{\text{SEE}}$ in \eqref{eq:SEE}, and $\mathbf{e}_{\phi_{X'}}, \mathbf{e}_{\phi_{Z'}}$ denote unit basis vectors representing the rotation angle components.

We can take the positive direction of the gradient as the ascent direction to maximize the objective. Then, the MA angles $\bm{\phi}_{\text{J}}^{(k)}[n]$ can be updated as
\begin{equation}
\small
\bm{\phi}_{\text{J}}^{(k+1)} = \mathcal{P}_{\mathcal{F}}\{\bm{\phi}_{\text{J}}^{(k)} + \alpha_k\nabla_{\mathbf{\phi}}f(\bm{\phi}_{\text{J}}^{(k)})\},
\end{equation}
where $\mathcal{P}_{\mathcal{F}}$ projects the solution onto the feasible set $\mathcal{F}$ which is determined by the constraint \eqref{eq:const3}. This approach ensures efficient convergence while maintaining physical constraints of the MA array \cite{ren20256d}. In addition, we use the backtracking method to get the appropriate step size that satisfies the Armijo–Goldstein condition:
\begin{equation}
\small
f(\bm{\phi}_{\text{J}}^{(k+1)}[n]) \geq f(\bm{\phi}_{\text{J}}^{(k)}[n]) + \kappa\cdot\alpha_k \lVert\nabla_{\mathbf{\phi}}f(\bm{\phi}_{\text{J}}^{(k)}[n])\rVert^2,
\end{equation}
where $\kappa\in\lbrack0,1\rbrack$ is a constant to achieve an adequate increase in the objective function with step size $\alpha_k$ determined by backtracking line search \cite{ren20256d}.

\vspace{-0.2cm}
\subsection{Optimizing Beamforming Vector Given Trajectory and Angles}
In this part, we optimize the objective with its outer product matrix $\mathbf{W}_{\text{J}}[n] = \mathbf{w}_{\text{J}}[n]\mathbf{w}_{\text{J}}^H[n]$ with the assumption that the BS employs maximum ratio transmission (MRT) beamforming towards the legitimate user. By substituting with $\mathbf{W}_{\text{J}}[n]$, the problem can be rewritten as:
\begin{subequations} \small
\begin{align}
\max_{\mathbf{W}_{\text{J}}[n] \succeq 0} \quad &\frac{\sum\limits_{n=1}^{N_{\text{step}}} \bar{r}_{\text{sec}}(\mathbf{W}_{\text{J}}[n])\Delta t}{ E_{\text{UAV}} + E_{\text{MA}} + P_{\text{J}}\sum\limits_{n=1}^{N_{\text{step}}}\text{Tr}(\mathbf{W}_{\text{J}}[n])\Delta t}, \tag{23} \label{eq:beam_obj} \\
\text{s.t.} \quad &\text{tr}(\mathbf{W}_{\text{J}}[n]) \leq 1, \quad \forall n, \label{eq:beam_const1} \\
                   &\text{rank}(\mathbf{W}_{\text{J}}[n]) = 1, \quad \forall n, \label{eq:beam_const2}
\end{align}
\end{subequations}

For effective optimization, we decompose the problem into $N_{\text{step}}$ sequential subproblems, where each subproblem aims to find the optimal $\mathbf{W}_{\text{J}}[n]$ at its corresponding time step.

After that, we relax the rank-1 constraint and transform the numerator of the objective into a concave lower bound by first-order Taylor expansion \cite{scutari2016parallel, zhong2018secure}. Therefore, the problem can be reformed to:

\begin{subequations} \label{eq:beam_sca} \small
\begin{align}
\max_{\mathbf{W}_{\text{J}}[n] \succeq 0} \quad &\frac{\omega_1 + \check{r}_{\text{sec}}(\mathbf{W}_{\text{J}}[n], \mathbf{W}_{\text{J}}^{(k)}[n])\Delta t}{ \omega_2 + P_{\text{J}}\cdot\text{Tr}(\mathbf{W}_{\text{J}}[n])\Delta t}, \tag{24} \label{eq:beam_sca_obj} \quad \text{s.t.} \quad \eqref{eq:beam_const1},
\end{align}
\end{subequations}
where $\omega_1$ and $\omega_2$ represent the remaining terms independent of $\mathbf{W}_{\text{J}}[n]$, and $\check{r}_{\text{sec}}$ is a concave lower bound of $\bar{r}_{\text{sec}}$.

Now, we again apply the Dinkelbach method to solve this fractional form optimization problem. After obtaining the optimal $\mathbf{W}_{\text{J}}[n]$, we can recover the MA beamforming vector through Gaussian randomization process.

To sum up, we solve for {$\mathbf{q}_{\text{J}}[n]$}, {$\bm{\phi}_{\text{J}}[n]$}, and {$\mathbf{w}_{\text{J}}[n]$} in an alternating manner, and consequently, our approach yields an efficient solution to the secrecy energy efficiency maximization problem. As the objective value is bounded and monotonically nondecreasing for every iteration, the convergence of the proposed optimization framework is guaranteed. The whole alternative process of our proposed framework is summarized in Algorithm I.

\begin{algorithm}[t]
\caption{Joint Optimization Algorithm for Secrecy Energy Efficiency Maximization}
{\footnotesize
\begin{algorithmic}[1]
\STATE \textbf{Input:} $N_{\text{B}}, N_{\text{MA}}, P_{\text{B}}, P_{\text{J}}, \mathbf{q}_{\text{B}}, \mathbf{q}_{\text{u}}, \mathbf{q}_{\text{e}}, \mathbf{q}_{\text{I}}, \mathbf{q}_{\text{F}}, \alpha_{\text{Bu}},\alpha_{\text{Be}}, \alpha_{\text{Ju}},\alpha_{\text{Je}},H_{\text{B}}, H_{\text{J}}$
\vspace{+0.1cm}
\STATE \quad\quad\quad $T_{\text{flight}}, N_{\text{step}}, V_{\text{max}}, P_0, P_1, A, S, r_{\text{drag}}, \rho, U_{\text{tip}}^{2}, v_0, P_{\text{base}}, \zeta,\xi, \epsilon,f$
\STATE \textbf{Output:} Optimized $\lbrace\mathbf{q}_{\text{J}}[n], \bm{\phi}_{\text{J}}[n], \mathbf{w}_{\text{J}}[n]\rbrace_{n=1}^{N_{\text{step}}}$
\STATE Initialize $\mathbf{q}_{\text{J}}^{(0)}[n]$, $\bm{\phi}_{\text{J}}^{(0)}[n]$, $\mathbf{w}_{\text{J}}^{(0)}[n]$
\STATE Set iteration index $k=0$ and convergence tolerance $\epsilon_{\text{th}}$
\REPEAT
    \STATE $k \leftarrow k + 1$
    \FOR{$n = 1$ to $N_{\text{step}}$}
    \STATE Update UAV trajectory $\mathbf{q}_{\text{J}}^{(k)}[n]$ by solving \eqref{eq:traj_sub_obj} with fixed $\mathbf{w}_{\text{J}}^{(k-1)}[n]$ and $\bm{\phi}_{\text{J}}^{(k-1)}[n]$
    \STATE Update MA orientation angles $\bm{\phi}_{\text{J}}^{(k)}[n]$ using two-phase approach with fixed $\mathbf{w}_{\text{J}}^{(k-1)}[n]$ and $\mathbf{q}_{\text{J}}^{(k)}[n]$
    \STATE Update beamforming vector $\mathbf{w}_{\text{J}}^{(k)}[n]$ by solving \eqref{eq:beam_obj} with fixed $\mathbf{q}_{\text{J}}^{(k)}[n]$ and $\bm{\phi}_{\text{J}}^{(k)}[n]$
    \STATE Calculate objective value $\eta_{\text{SEE}}^{(k)}$
    \IF{$|\eta_{\text{SEE}}^{(k)} - \eta_{\text{SEE}}^{(k-1)}| < \epsilon_{\text{th}}$}
        \STATE Break
    \ENDIF
    \ENDFOR
\UNTIL{convergence or maximum iterations reached}
\RETURN $\lbrace\mathbf{q}_{\text{J}}^{\ast}[n], \bm{\phi}_{\text{J}}^{\ast}[n], \mathbf{w}_{\text{J}}^{\ast}[n]\rbrace_{n=1}^{N_{\text{step}}}$
\end{algorithmic}
}
\end{algorithm}

\vspace{-0.3cm}
\section{Numerical Results}
In this section, we demonstrate the effectiveness of our proposed optimization framework through several simulations. The system consists of a BS with $N_{\text{B}}=4$ transmit antennas, a single-antenna user, and a potential eavesdropper. The UAV operates at $H_{\text{J}}=50$m with maximum velocity $V_{\text{max}}=15$m/s. The rest of parameters are provided in Table \ref{table:1} \cite{yang2019energy}. The simulations show the performance of energy consumption and efficiency over iteration number and UAV trajectory for four different approaches: proposed optimal approach, an Eve-oriented approach where the MA array always points towards the eve, a direct path case where the UAV follows a straight trajectory, and a conventional fixed-antenna approach.

\begin{figure}[t]
\vspace{-0.3cm}
\centering
\subfloat[Energy consumption]{
\includegraphics[width=0.465\columnwidth]{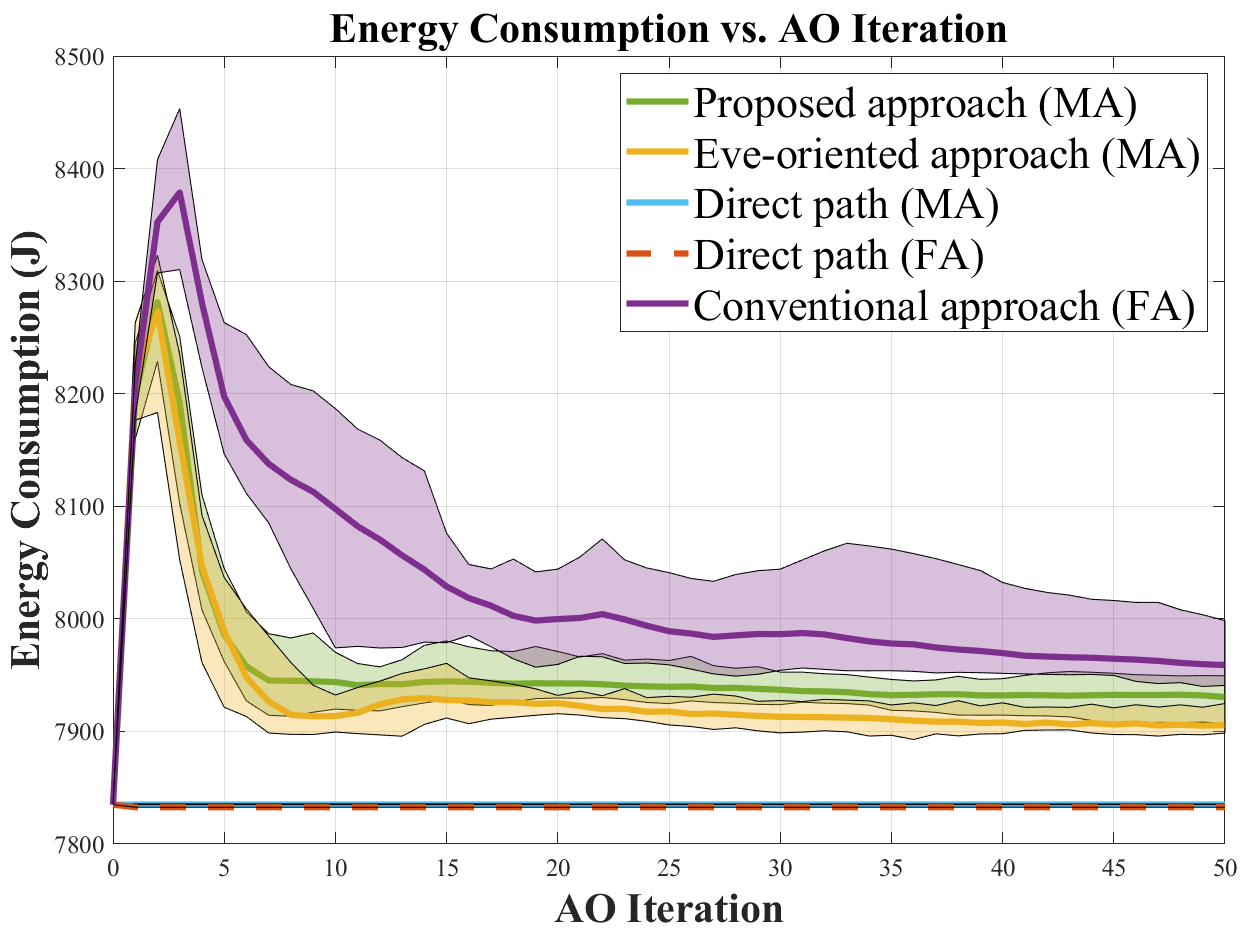}
\label{fig:secrecy_energy}
}
\hfill
\subfloat[Energy efficiency]{
\includegraphics[width=0.465\columnwidth]{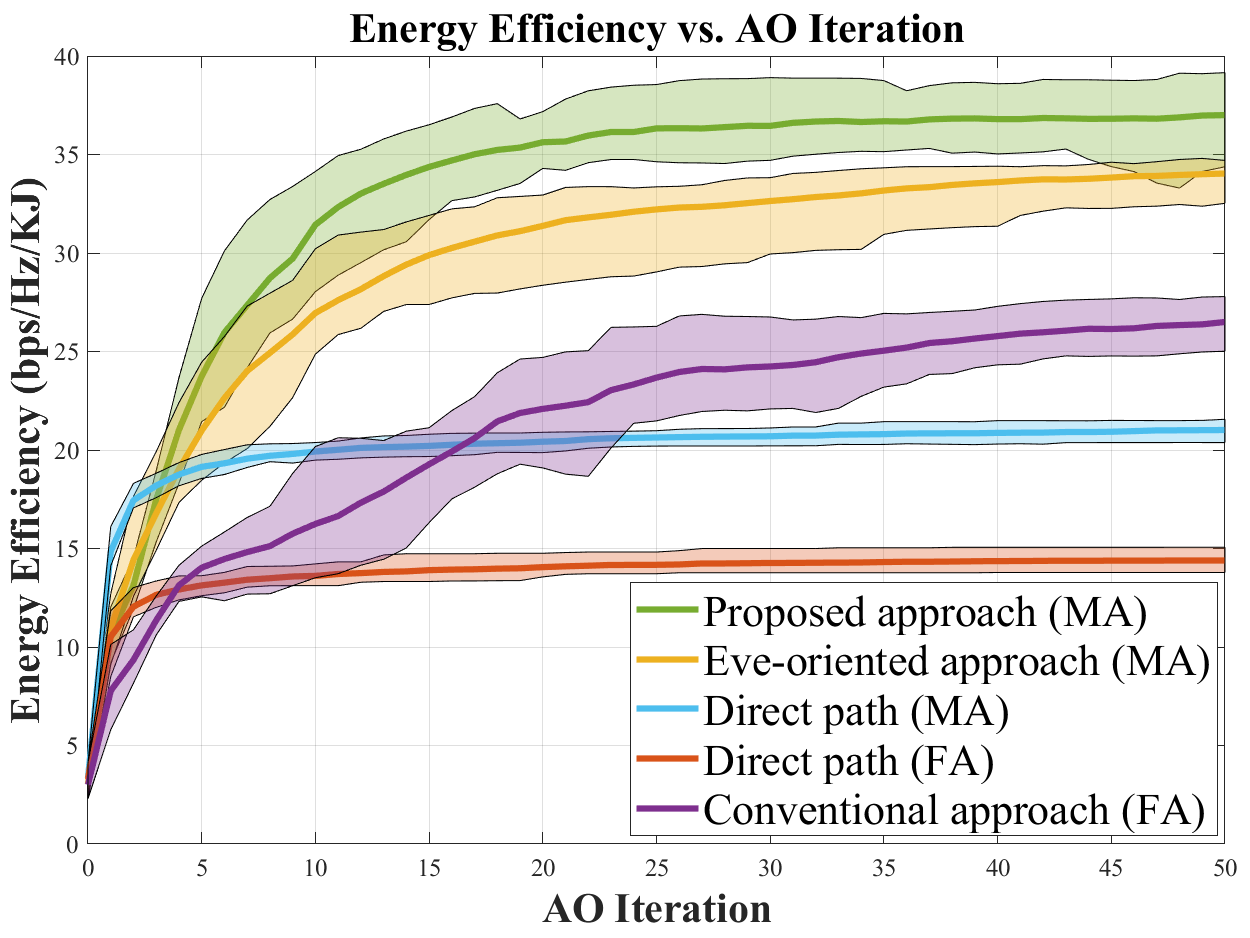}
\label{fig:energy_efficiency}
}
\vspace{-0.1cm}
\caption{Performance comparison of different approaches}
\label{fig:performance}
\end{figure}

The results reveal several key insights regarding the performance of different approaches. Notably, the use of movable antennas consistently demonstrates lower energy consumption compared to fixed-antenna systems as shown in Fig. \ref{fig:performance}, regardless of the optimization strategy employed. Even though the direct path achieves minimal energy consumption, our proposed method maintains similar levels with enhanced security benefits. In Fig. \ref{fig:energy_efficiency}, it is remarkable that the proposed method achieves a better performance than any other approaches. While our proposed approach maintains energy consumption levels similar to the Eve-oriented method, it achieves approximately 10\% higher energy efficiency. Moreover, our algorithm demonstrates around 40\% improvement in efficiency, when compared to systems with fixed antennas. This improvement stems from the optimal adjustment of MA orientation angles, which enables the creation of strong beams towards the eve while forming nulls in the legitimate user's direction, all while minimizing energy consumption.

\begin{figure}[t]
\vspace{-0.2cm}
\centering
\includegraphics[width=0.4\textwidth]{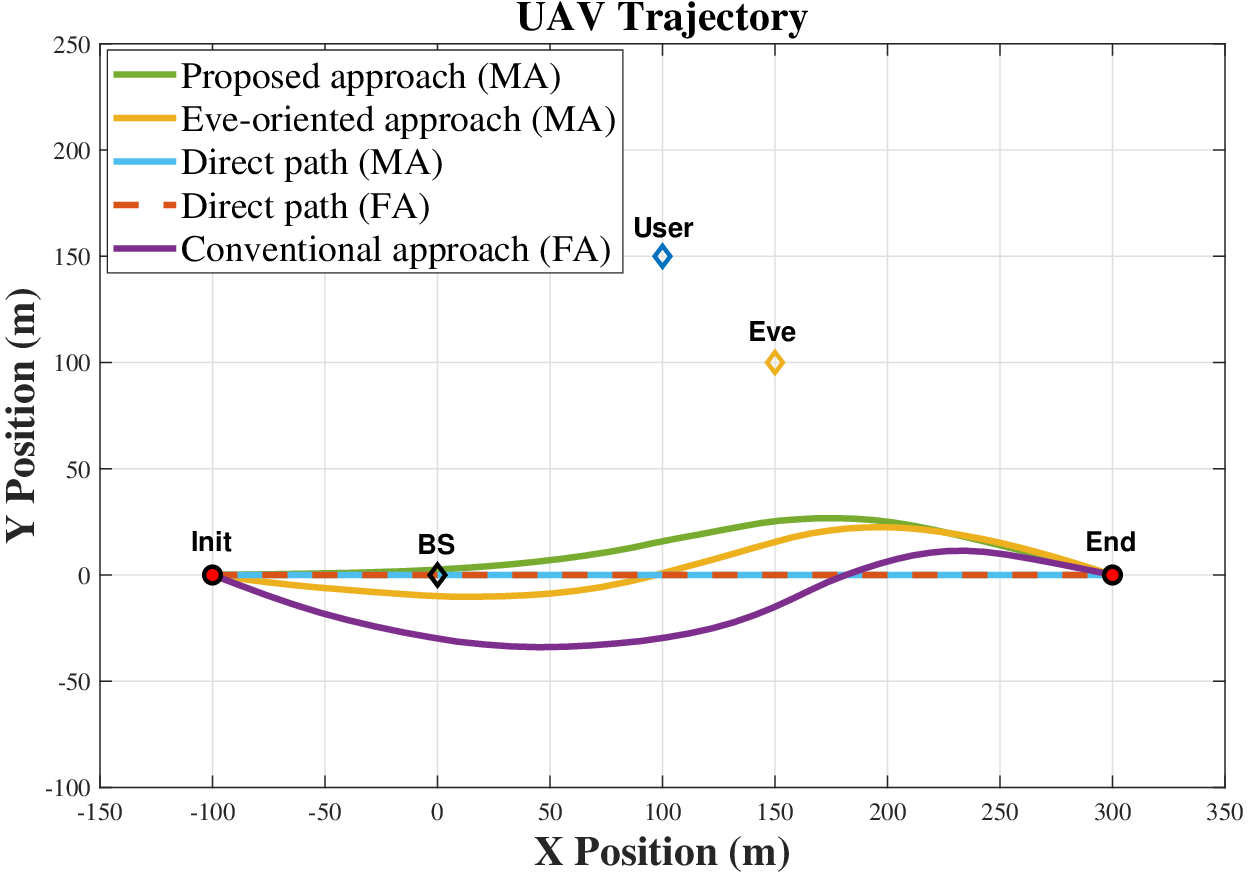}
\vspace{-0.1cm}
\caption{Optimized UAV trajectory and node locations}
\label{fig:trajectory}
\end{figure}

The trajectory analysis in Fig. \ref{fig:trajectory} provides further insights into the behavior of different approaches. The fixed-antenna system exhibits a more pronounced curved trajectory, where the UAV tends to move with larger curvature around the legitimate user to maximize secrecy performance. This behavior results in longer flight paths and increased energy consumption. In contrast, our MA-based approach demonstrates a more efficient trajectory. By leveraging the additional degrees of freedom provided by the MA, the UAV can maintain a shorter path while positioning itself closer to the eve to maximize jamming effectiveness. This strategic positioning and flexible beamforming capabilities of the MA array, enabled by our proposed joint optimization framework, effectively leverages additional degrees of freedom to achieve optimal channel conditions, leading to superior secrecy performance and energy efficiency compared to conventional approaches.

\begin{table}[t!]
\vspace{-0.1cm}
\raggedright
\caption{Simulation settings}
\vspace{-0.3cm}
\label{table:1}
\begin{center}
\footnotesize
\scalebox{0.8}{
\begin{tabular}{|c|c||c|c|}
\hline
Parameter & Value & Parameter & Value \\\hline
$N_{\text{B}}, N_{\text{MA}}$ & 4, 16 & $N_{\text{step}}$ & 40 step \\\hline
$H_{\text{B}},H_{\text{J}}$ & 12.5, 50 (m) & $\mathbf{q}_{\text{B}}$ & (0,0,$H_{\text{B}}$) \\\hline
$P_{\text{B}}, P_{\text{J}}$ & 100, 10 (W) & $\mathbf{q}_{\text{u}}$ & (100,150,0) \\\hline
$\alpha_{\text{B},u},\alpha_{\text{B},e}$ & 3.5 & $\mathbf{q}_{\text{e}}$ & (150,100,0) \\\hline
$\alpha_{\text{J},u},\alpha_{\text{J},e}$ & 2.8 & $\mathbf{q}_{\text{I}}$ & (-100,0,$H_{\text{J}}$) \\\hline
$\sigma_u^2,\sigma_e^2$ & -114 dBm & $\mathbf{q}_{\text{F}}$ & (300,0,$H_{\text{J}}$) \\\hline
$f$ & 28GHz & $\epsilon$ & 50 m \\\hline
$T_{\text{flight}}$ & 40 s & $v_{\text{max}}, v_0$ & 15, 2.5669 (m/s) \\\hline
$P_{\text{base}}$ & 2 W & $\zeta,\xi$ & 0.05, 0.03 W/rad \\\hline
$P_0, P_1$ & 125.4, 200 & $(A, S)$ & (0.79, 0.05)\\\hline
$(r_{\text{drag}}, \rho)$ & (0.6, 1.225) & $U_{\text{tip}}^{2}$ & 8100 \\\hline
\end{tabular}
}
\end{center}
\vspace{-0.3cm}
\end{table}

\vspace{-0.3cm}
\section{Conclusions}
This work demonstrated that UAVs equipped with movable-antenna (MA) arrays can significantly improve secure wireless communications through a joint optimization of trajectory, beamforming, and antenna orientation. By leveraging both UAV mobility and MA flexibility, the proposed framework achieves higher secrecy energy efficiency compared to conventional methods. The joint optimization of system parameters—from UAV paths to antenna configurations—offers a promising direction for next-generation secure wireless systems, particularly in scenarios demanding strong physical layer security and efficient resource use.

%

\bibliographystyle{IEEEtran}
\bibliography{UAV_MA_refer}

\end{document}